\documentclass[10pt,twocolumn,preprintnumbers,superscriptaddress,nofootinbib,aps,prl]{revtex4-2}

\usepackage{graphicx}
\usepackage{amsmath}
\usepackage{srcltx}
\usepackage[utf8]{inputenc}
\usepackage[colorlinks, pdfusetitle]{hyperref}
\usepackage{times}
\usepackage{amssymb}
\usepackage{color}

\newcommand{\vev}[1]{\langle {#1} \rangle}
\newcommand{\lsim}{\lesssim}
\newcommand{\gsim}{\gtrsim}

\newcommand{\eq}[1]{Eq.~(\ref{#1})}

\newcommand{\ord}[1]{\mathcal{O}{(#1)}}
\newcommand{\beq}{\begin{equation}}
\newcommand{\eeq}{\end{equation}}

\newcommand{\mP}{M_{\rm P}}

\newcommand{\orcid}[1]{\href{https://orcid.org/#1}{#1}}

\newcommand{\prlsection}[2]{{\it\textbf{#1}{#2}}---}

\begin{document}
\preprint{FERMILAB-PUB-25-0683-T}

\pagestyle{plain}

\title{\boldmath Cosmic Overture Echoed in a Stellar Final Act:\\ Implications of Nanohertz Gravitational Waves for Core Collapse Supernova Neutrinos}

\author{Hooman Davoudiasl}
\email{hooman@bnl.gov}
\thanks{\orcid{0000-0003-3484-911X}}
\affiliation{High Energy Theory Group, Physics Department \\ Brookhaven National Laboratory, Upton, NY 11973, USA}

\author{Peter B.~Denton}
\email{pdenton@bnl.gov}
\thanks{\orcid{0000-0002-5209-872X}}
\affiliation{High Energy Theory Group, Physics Department \\ Brookhaven National Laboratory, Upton, NY 11973, USA}

\author{Anna M.~Suliga}
\email{a.suliga@nyu.edu}
\thanks{\orcid{0000-0002-8354-012X}}
\affiliation{Center for Cosmology and Particle Physics, New York University, New York, NY 10003, USA}

\begin{abstract}
The growing evidence for nanohertz gravitational waves, from NANOGrav and other observations, may be pointing to a cosmological first-order phase transition at temperatures of $\ord{10-100}\;\mathrm{MeV}$.
Such an interpretation requires beyond the Standard Model dynamics in this energy range.
If so, it may well be the case that galactic core-collapse supernova explosions would produce the key components related to the first-order phase transition, leaving detectable imprints on the spectrum of neutrinos emitted in the initial few seconds of the collapse.
This might provide further evidence in support of the early universe interpretation of the nanohertz gravitational wave signal.
The scenario proposed here is also suggestive of a low-mass seesaw mechanism to explain neutrino masses.
We outline the prospects for future observations of Galactic supernovae to uncover the signals of this scenario, with further confirmation from future pulsar timing array measurements of primordial nanohertz gravitational waves.
\end{abstract}

\maketitle

\prlsection{Introduction}{.}
The outstanding puzzles of particle physics and cosmology provide strong motivation for new physics.
While the physics resolving these puzzles may have limited interactions with the Standard Model (SM) particles, since all forms of energy couple to gravity, one may hope that gravitational data could shed light on the nature and scales associated with new physics.
In particular, new mass scales may arise from cosmological phase transitions leading to primordial gravitational waves (GWs) that encode features of the underlying physics.
For these reasons, detection of the background GWs from the early Universe can offer clues to hidden sectors that are otherwise difficult to access and probe. 

Recently, a number of pulsar timing arrays reported evidence for stochastic GWs in the nanohertz regime, notably NANOGrav~\cite{NANOGrav:2023gor}, adding to earlier hints \cite{Manchester:2012za, EPTA:2016ndq, Perera:2019sca, NANOGrav:2020bcs, EPTA:2021crs, Goncharov:2021oub, Antoniadis:2022pcn}. Such a background can be expected to arise from a number of sources.
One such candidate is of astrophysical origin, {\it i.e.}~inspiraling supermassive black hole binaries (SMBHBs) leading to a low frequency background of GWs~\cite{Rajagopal:1994zj, Jaffe:2002rt, Wyithe:2002ep, Enoki:2004ew, Sesana:2012ak, McWilliams:2012an, Burke-Spolaor:2018bvk}. This source is generally expected to contribute to the reported signal, given that supermassive black holes have been observed and are expected to be a common feature of galactic centers~\cite{Volonteri:2021sfo}. However, the data may be showing signs of an additional component that prefers a primordial source, in particular an early Universe first-order phase transition (FOPT)~\cite{Witten:1984rs, Hogan:1986qda, Kosowsky:1992rz, Kamionkowski:1993fg} at a temperature 
$T_*\sim \ord{10-100}\;\text{MeV}$
\cite{NANOGrav:2023hvm}.
The inferred parameters of the FOPT depend on whether the GWs are generated by collisions of true vacuum bubbles or sound waves due to the motion of primordial plasma (see, {\it e.g.}, Ref.~\cite{Athron:2023xlk} for a recent review of relevant physics).

In a FOPT, it is expected that bubble collisions are dominant when the wall achieves runaway behavior, indicating weak interactions with the plasma, while significant coupling between the wall and the plasma would lead to conversion of the wall kinetic energy to sound waves. This is a model dependent question and the current data is not yet precise enough to shed light on it.
In general, the fit to the GW data is also affected by whether one includes contributions to the GW signal from SMBHBs, which is subject to astrophysical uncertainties.
The GW data does not provide an indication of the particle nature of the FOPT, but new physics at the implied mass scales of $\sim$ 10-100~MeV is generically assumed to have suppressed interactions with electrically charged SM states.
Nonetheless, it may well have non-negligible couplings to neutrinos, on which we focus here.

In this work, we will assume that the nanohertz GW data is generated by an early Universe FOPT which is realized by a scalar field that is thermalized via couplings to neutrinos, as these generally have the weakest constraints and are consistent with all relevant particle physics and cosmology data.
We will mostly follow the discussion provided by the NANOGrav collaboration for the inferred parameter space of the FOPT \cite{NANOGrav:2023hvm}.
There have been several more detailed models realized to implement FOPTs to explain the NANOGrav data in a variety of ways \cite{Guo:2023hyp, Salvio:2023blb, Salvio:2023ynn, Costa:2025csj, Chatrchyan:2025wop, Goncalves:2025uwh, Correia:2025qif, Lyu:2025jge, Barenboim:2025jai}.
We take the fairly model independent view that some transition characterized by a temperature $T_*$ in the range $\sim$ 10-100~MeV can explain  the data.
Our main observation is that such a temperature range has significant overlap with those of a core-collapse supernova (CCSN) during the first few seconds after the core bounce~\cite{Bruenn:2012mj, Takiwaki:2013cqa, OConnor:2018sti, Mirizzi:2015eza, Janka:2016fox, Mezzacappa:2020oyq, Burrows:2020qrp, Raffelt:2025wty}.  Hence, we find it plausible that the conditions during the CCSN can be sufficiently close to that of the cosmic epoch corresponding to the FOPT that generated the nanohertz GWs.  This would lead to the production of the key physical states associated with the early Universe FOPT in the initial seconds following the progenitor collapse, for a broad class of models under fairly generic assumptions.
Moreover, our scenario naturally leads to an explanation of neutrino masses via a low scale seesaw mechanism \cite{Minkowski:1977sc, Yanagida:1979as, Gell-Mann:1979vob, Yanagida:1980xy, Mohapatra:1979ia, Gelmini:1980re, Schechter:1981cv}, see also \cite{Bi:2025wyp,Jana:2025vyb,Samanta:2020cdk,Datta:2020bht,Datta:2023vbs,Chianese:2024gee,Chianese:2025mll} for other scenarios connecting GWs and seesaws.

We also point out that the early Universe setting is expected to be governed by a homogeneous plasma which, for our typical temperature range, has a low baryon density. This is in contrast to the baryon dense supernova environment, which also features significant inhomogeneities. However, we focus on the couplings of the new particles of the phase transition to neutrinos. Given that the neutrino population in a supernova is also characterized by a largely homogeneous relativistic fluid, we expect that the key features of the assumed early Universe dynamics would be {\it re-enacted} in the initial stages of a sufficiently hot supernova.

In the scenario we entertain, during the last hundreds of milliseconds before the shock runaway, the core of the CCSN heats up and populates the hidden sector possibly leading to a thermal restoration of the false vacuum at and around the stellar core. The potential restoration of the false vacuum followed by the phase transition back to the true vacuum, however, is not strictly necessary for the phenomenology discussed in this paper.
Regardless, as the CCSN starts to cool the scalars produced in the CCSN lead to detectable modifications of the neutrino signal.
Moreover, the coupling of the scalar field can generate the correct size of mass for the active neutrinos, however this requires massive right handed neutrinos which can naturally lead to a low-mass type-I seesaw.  
In the next section we will discuss the connection between the FOPT in the early Universe and a CCSN.

\prlsection{Early Universe Dynamics}{.}
As stated before, we do not adopt a particular model.
However, we simply assume a potential of the form
\beq
V(\phi) = -\frac{\mu_\phi^2}{2} \, \phi^2 + \frac{\bar\mu}{3!} \,\phi^3 + \frac{\lambda}{4!}\,\phi^4\,,
\label{Vphi}
\eeq
where $\mu_\phi^2 >0$ is the tachyon mass squared parameter, $\bar\mu$ is a mass scale that leads to a FOPT through formation of a barrier between the false and true vacuum during the transition, and {$\lambda \sim \mathcal{O}(1)$} is the dimensionless quartic coupling.
The self-coupling in this potential will lead to a restoration of the false vacuum via thermal effects at temperatures above the critical temperature $\mathcal O(10$ MeV$)$.
This potential has a softly broken $\mathbb Z_2$ symmetry through the cubic term, which we assume to be the only source of this  breaking. 

For temperatures of $\ord{10~\text{MeV}}$ in the early Universe, the SM degrees of freedom are 6 neutrinos and antineutrinos, electron, positron, and the photon.  This yields $g_*(SM) = 10.75$ degrees of freedom from the SM.  To implement the transition, $\phi$ is taken to be a thermalized degree of freedom which would generically require some level of interaction between the SM and the new physics.  Since at the relevant epoch the density of baryons is $n_B (T)\sim 10^{-10} T^3$, a simple estimate for establishing thermal equilibrium is given by
\beq
\frac{g_B^2 n_B(T_*)}{m_n^2} \gsim {\cal H}(T_*)\,,
\label{th-eq}
\eeq       
where $g_B$ represents a typical coupling between the hidden sector and baryons, and ${\cal H}(T)\sim g_*^{1/2} T^2/\mP$ is the Hubble rate at temperature $T$; $\mP\approx 1.2 \times 10^{19}$~GeV is the Planck mass.  For $T_*\lsim 50$~MeV, as a value typical of our scenario, we find that  $g_B\gsim 10^{-4}$ is required, which is generally ruled out for scalars lighter than $\sim 100$~MeV \cite{Knapen:2017xzo}.  

Given that electrons are abundant during the epoch of interest here, one could achieve thermalization more readily through them.  A simple estimate of thermalization through decay and inverse decay processes $e^+ e^- \leftrightarrow \phi$ suggests that the electron coupling needed is $\gsim 10^{-9}$, which is nearly ruled out by cosmological constraints~\cite{Knapen:2017xzo} for $m_\phi \gsim 10$~MeV.\footnote{The constraints on the scalars coupling to electrons from CCSN might be even stronger for the pseudoscalar example~\cite{Fiorillo:2025sln}.}  We note that the coupling of a boson (like our $\phi$ scalar) of mass $\sim \ord{\text{10~MeV}}$ to muons is similarly constrained to be $\lsim 10^{-9}$, but due to CCSN energy loss considerations \cite{Bollig:2020xdr,Croon:2020lrf}.  Since the muon population is somewhat Boltzmann suppressed at temperatures of interest in the early Universe, we thus conclude that they are not expected to lead to a thermalized $\phi$ population either.         

In light of the above considerations,
we will assume that $\phi$ is thermalized by their interactions with thermally abundant neutrinos.
These interactions with neutrinos are less constrained than those with other SM states and thus are safely allowed by other data.
Then, since a CCSN results in a large thermal population of neutrinos and antineutrinos, we expect that the $\phi$'s would be similarly populated in the stellar collapse and the consequent explosion.

Under minimal assumptions, we let the interaction of the scalar $\phi$ with neutrinos be of the form
\beq
\frac{{\tilde H}^*  \,\bar L\,\phi\,\nu_R}{M} 
+ \text{\small H.C.}
\to 
y \, \phi \bar \nu_L \nu_R + \text{\small H.C.}
\label{Yukawa}
\eeq
with, $H$ the SM Higgs doublet, $L$ a lepton doublet, $\nu_L$ denoting an active neutrino, and $\nu_R$ a right-handed SM singlet.
In \eq{Yukawa}, $y\equiv \vev{H}/M$, where $M$ is an ultraviolet scale and $y$ will have some flavor structure.  
We note that the above interaction may be sufficient for a Dirac mass.  However, $\nu_R$ will also be brought into thermal equilibrium once $\phi$ is thermalized.  We would then end up with too many relativistic degrees of freedom at Big Bang Nucleosynthesis (BBN) \cite{Schoneberg:2024ifp} and later, as in this case $\nu_R$ is massless and cannot decay.  Hence, we are naturally led to a seesaw mechanism with Majorana $\nu_R$ states of mass $m_R$.  As we want $\nu_R \to \nu_L \phi$ to be allowed -- in order to deplete the $\nu_R$  population before BBN -- we require $m_R > m_\phi$, where $m_\phi$ is the mass of $\phi$.  To highlight the main physics picture, we will simply assume that $m_\phi \lsim \vev{\phi}$, which can be realized in a generic scalar potential.  We roughly have 
\beq
\Gamma(\nu_R \to \nu_L \phi,\, \bar\nu_L \phi)\sim \frac{y^2\,m_R} {16\pi} \ .
\label{GammanuR}
\eeq
Requiring that $\nu_R$ decay before BBN, at $T\lsim 4$~MeV \cite{Hannestad:2004px}, would yield $y\gsim 10^{-10}$, for $m_R\gsim 10$~MeV.  

We now implement the seesaw and determine the viable ranges for the masses and couplings.
Given the preceding discussion, in our scenario we typically expect $m_R > T_*$, making $\nu_R$ a heavy degree of freedom, compared to the energy scales of the FOPT $\ord{10~{\rm MeV}}$.  This lends itself well to a seesaw picture of neutrino masses $m_\nu$, with   
\beq
m_\nu \sim \frac{y^2 \vev{\phi}^2}{m_R}\sim0.1{\rm\ eV}\,.
\label{mnu}
\eeq
Let us set $\vev{\phi}\sim 30$~MeV, as may be typical of  our setup, for the sake of concreteness.  Neutrino masses are then valid for a range of $y$ and $m_R$ so long as
\beq
\frac{y^2}{m_R}\simeq10^{-10}\;\ \text{MeV}^{-1}\,.  
\label{eq:seesaw}
\eeq
In our scenario,  we then have $y\gtrsim3\times 10^{-5}$ which comes from the requirement that $m_R \gsim 10$~MeV (on-shell $\nu_R$ decay into $\phi$).
We note that low energy $\phi$ interactions 
\beq
\xi_\nu \phi\,\bar \nu_L^c \nu_L \ ,
\label{phinunu}
\eeq
with active neutrinos lead to Majorana masses for SM neutrinos and {\it lepton number violation} (LNV) mediated by $\phi$.  In our scenario, the coupling (matrix) $\xi_\nu$ is set by the typical vev  $\vev{\phi}\sim 30$~MeV via $\xi_\nu \sim m_\nu/\vev{\phi}\sim 10^{-9}$.
Such coupling is however too weak to trap the scalar particles inside the core of the CCSN leading to excessive cooling of the supernova core (see, for example, Refs.~\cite{Heurtier:2016otg, Akita:2022etk, Fiorillo:2022cdq}).  This constraint naturally leads us to the regime $m_R\lsim 100$~MeV, so that $\phi$ scattering $\nu_L \phi \to \nu_R$ is allowed in CCSN and followed by prompt $\nu_R\to \phi\, \nu_L$ and thus and upper limit on $y$ of $10^{-4}$ to satisfy the seesaw constraint (\ref{eq:seesaw}).
The allowed range of couplings from the above considerations is $3\times10^{-5}\lsim y\lsim 10^{-4}$ set by the seesaw\footnote{The entire $y$ range of interest is safe from the laboratory bounds on meson decays and $Z$-width measurements, which for $m_\phi \lesssim 100\;$MeV restrict the coupling to be smaller than $y \lesssim 4\times 10^{-3}$~\cite{Pang:1973rxr, Bilenky:1992xn, Laha:2013xua, Pasquini:2015fjv, Berryman:2018ogk, Dev:2024ygx} (assuming that $\nu_R$ is sufficiently light to allow the relevant decay processes, but $\nu_R$ must still be heavier than $\phi$ so it can decay).}.
These processes are governed by the relatively larger coupling $y$ and would not allow $\phi$ to free-stream out of the CCSN core.  As $\nu_R$ are Majorana particles, the above processes lead to {\it rapid} LNV relative to the weak interaction.
The LNV timescale at our benchmark point is $\ll\mathcal O($ns$)$ while the weak interaction is $\gtrsim\mathcal O($ns$)$.
We show benchmark values of our parameters in Table \ref{tab:parameters}.

\begin{table}
\caption{Benchmark values for the parameters used in the paper.}
\label{tab:parameters}
\begin{tabular}{c|c|c|c}
$\vev{\phi}$ & $m_\phi$ & $y$ & $m_R$ \\\hline
30 MeV & 10 MeV & $10^{-4}$ & 100 MeV
\end{tabular}
\end{table}

Let us estimate the range of parameters for which the interaction in \eq{Yukawa} can lead to thermalization of $\phi$ and a consequent FOPT in the early Universe.  For $m_R\gsim T \gsim m_\phi$, we may write the rate for $\nu \nu \to \phi \phi$ as 
\beq
\Gamma(\nu\nu \to \phi \phi) \sim \frac{y^4 \,T^3}{m_R^2}\sim \frac{m_\nu^2\,T^3}{\vev{\phi}^4}\,,
\label{Gam2nu2phi}
\eeq
where we have used \eq{mnu}.  Demanding that the rate in \eq{Gam2nu2phi} be larger than ${\cal H}(T)$, we find that there is plenty of space available to achieve thermalization all the way down to $T\lsim T_*$, and hence $\phi$ will be thermalized for the range of parameters considered.

%%%%%%%%%%%%%
\begin{figure}
\centering
\includegraphics[width=0.98\columnwidth]{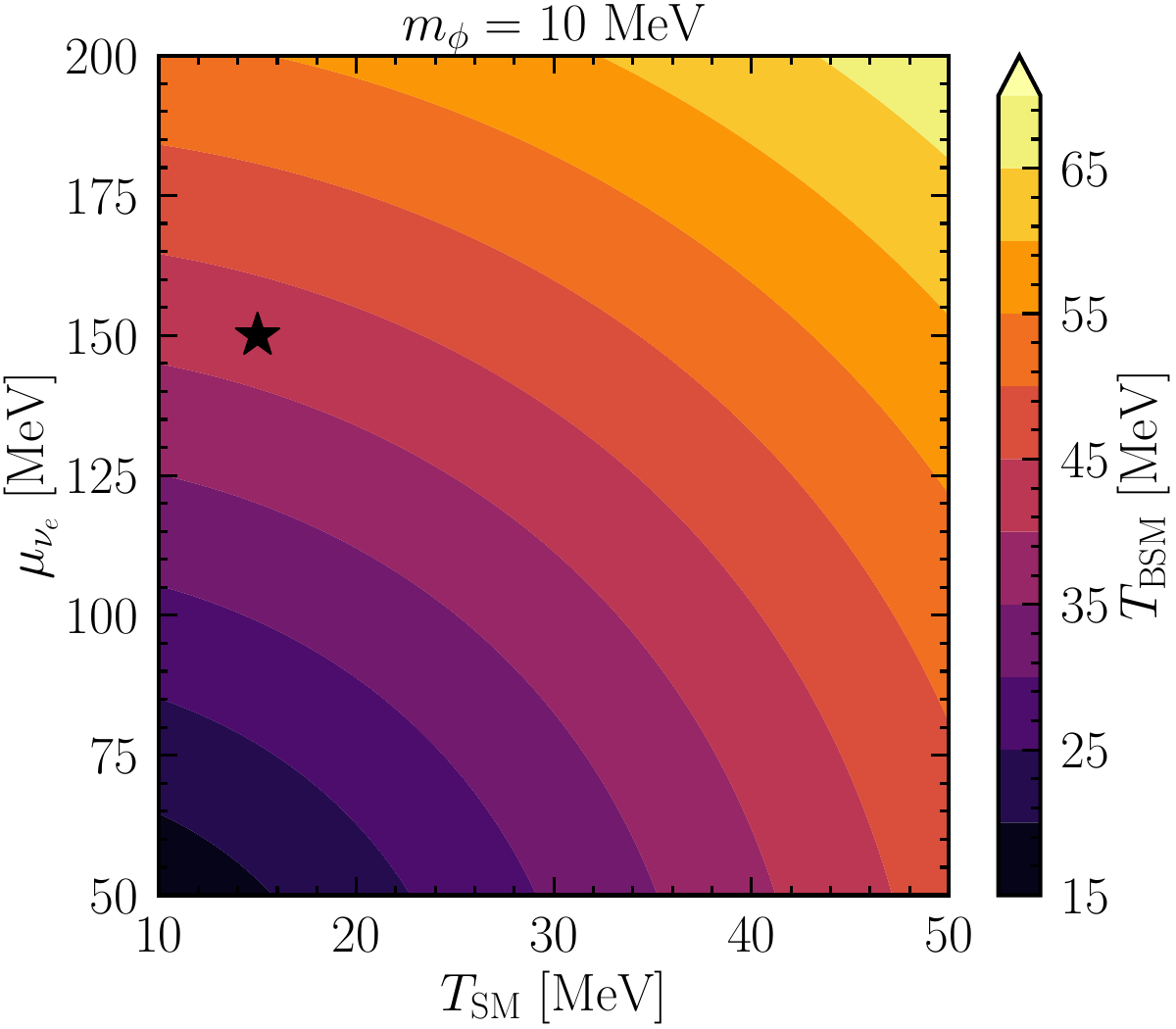}
\caption{The temperature ($T_\mathrm{BSM}$) after the all six neutrino flavors, two sterile neutrinos at $m_R=100$ MeV, and scalar equilibration as a function of the SM CCSN temperature ($T_\mathrm{SM}$) and chemical potential of electron neutrinos ($\mu_{\nu_e}$).
The black star depicts a representative value for the conditions just after the core bounce in SM CCSN core. Note that the final temperature $T_\mathrm{BSM}$ achieved by the equilibration of the massive scalars with neutrinos will be reduced if the scalars undergo the FOPT.}
\label{fig:phi-equlibration}
\end{figure}
%%%%%%%%%%%%%

Implicit in the above is the assumption that $\vev{\phi}$ sources the majority of the mass for SM neutrinos, while the dimension-4 Yukawa coupling $H\bar \nu_L \nu_R$ is small or absent.  This could be the result of charging $\nu_R$ and $\phi$ under a $\mathbb Z_2$ symmetry which would still allow a bare Majorana mass term $m_R {\bar \nu_R^c} \nu_R$ to establish a seesaw for $m_\nu$.  As mentioned before, this $\mathbb Z_2$ symmetry is assumed to be only softly broken by the cubic interaction in \eq{Vphi}, which does not generate the ignored dimension-4 Yukawa coupling.  Hence, we can justify the smallness of larger Dirac mass terms that do not depend on $\vev{\phi}\neq 0$.    

\prlsection{Supernova Dynamics}{.}
We now consider the impact of the scalar field, sourcing the FOPT in the early Universe, on the neutrino signal in a CCSN.  One is the formation of a {\it hidden-sphere} composed of $\phi$'s and $\nu_R$'s \cite{Davoudiasl:2005fd}.
This hidden-sphere is analogous to the neutrino-sphere, where all flavors of $\nu$ and $\bar \nu$ form a thermal population that is trapped, due to its high density, in the initial seconds after the collapse.
In fact, we expect that the two spheres should roughly overlap and have the same temperature profiles.
These spheres are likely overlapping due to the fact that neutrinos are trapped inside the neutrino-sphere. Therefore, as neutrinos are trapped the thermalization of them with the $\phi$ field, which happens on timescales faster than weak interactions as mentioned earlier, will take place in the same region.
Due to the Boltzmann suppression of the $\nu_R$ population, with $m_R\sim100$ MeV, relative to the $\phi$ population, the $\nu_R$ contribution to the CCSN dynamics will be subdominant and thus we will ignore the $\nu_R$ population in the following discussion of observable signatures.

As with the usual thermal neutrinos, we expect the MeV scale $\phi$ to be radiated from the surface of the hidden-sphere.  
The emitted bosons would then decay into neutrinos in a detectably time-dependent signal.  We note the decay of $\phi$ will typically happen well outside the neutrino/hidden sphere.  This is due to the relatively small coupling $\xi_\nu\sim 10^{-9}$ of $\phi$ with neutrinos from Eq.~(\ref{phinunu}), which leads to a proper lifetime 
\beq
\tau_\phi \sim \left(\frac{\xi_\nu^2 \, m_\phi}{16 \pi}\right)^{-1}\,.
\label{tauphi}
\eeq
For our benchmark parameters the above yields $\tau_\phi\sim 0.003$~s; in the star frame this acquires a boost factor of $\sim3$ for a lifetime of $\sim0.01$~s.  Given that $\phi$ bosons would be emitted at velocities of order the speed of light, we then expect them to decay $\sim3\times10^3~\text{km}$ away from the boundary of the neutrino/hidden sphere which means that the neutrinos would not rethermalize.

However, as noted in Ref.~\cite{Davoudiasl:2005fd}, we expect the emitted neutrinos to inherit the Bose-Einstein distribution of the parent $\phi$ bosons.
After integrating over solid angle, this modifies the spectrum from a Bose-Einstein distribution, but still remains distinct from a (pinched) Fermi-Dirac distribution.
These neutrinos typically carry about half the energy of neutrinos decoupling from the core, which is potentially distinguishable from the standard pinched Fermi-Dirac distribution \cite{Keil:2002in}.
The $\phi$'s will be populated in the CCSN via Eq.~\eqref{Gam2nu2phi} on timescales that are fast, $\mathcal O($ms$)$, on CCSN  neutrino signal variability timescales which are $\mathcal O(0.1$s$)$, since they are thermalized in the early Universe for $T\gsim 10$~MeV.

Another important effect is that the presence of the scalar field will also drive the neutrino chemical potentials to zero due to the rapid LNV.
We have verified this for our scenario consistent with the result in Ref.~\cite{Suliga:2024nng} that explicitly calculates the impact of the fast LNV neutrino interaction on the $\beta$-equilibrium in the infall phase of the CCSN. (See also Refs.~\cite{Kolb:1981mc, Fuller:1988ega} that assume such an effect and discuss~\cite{Kolb:1981mc} and simulate~\cite{Fuller:1988ega} consequences for the CCSN evolution; Refs.~\cite{Kolb:1987qy, Shalgar:2019rqe, Reddy:2021rln, Chang:2022aas, Fiorillo:2023ytr, Fiorillo:2023cas, Wang:2025qap, Ivanez-Ballesteros:2024nws} provide some other studies of scalar mediated neutrino self-interactions~\cite{Bialynicka-Birula:1964ddi} in CCSNe.) 
This, in turn, will typically increase the maximum temperature of the CCSN core.
While there is some variation at the $\mathcal O(10\%)$ level on the expected temperature of a CCSN from simulations, see {\it e.g.} Ref.~\cite{OConnor:2018sti}, the anticipated change should be much larger leading to a potentially observable signal in the neutrinos detected from a future galactic CCSN observation. In addition, the electron fraction of the core is expected to decrease following the vanishing of neutrino chemical potentials~\cite{Suliga:2024nng, Fuller:1988ega}.

To understand the impact of the new physics on the temperature of the CCSN, we solve the energy conservation equation:
\begin{multline}
\sum_{\alpha\in\{e,\mu,\tau\}}\rho_{\nu_\alpha}(T_{\rm SM},\mu_{\nu_\alpha})=\\\sum_{\alpha\in\{e,\mu,\tau\}}\rho_{\nu_\alpha}(T_{\rm BSM},0)+\rho_\phi(T_{\rm BSM},0)+\rho_{\nu_R}(T_{\rm BSM},0)\,,
\end{multline}
where $\rho$ is the energy density and the sum includes both neutrinos and antineutrinos.
Fig.~\ref{fig:phi-equlibration} shows the effect of equilibration between all six neutrino flavors, \emph{i.e.}~$\nu_e, \bar\nu_e, \nu_\mu, \bar \nu_\mu, \nu_\tau, \bar\nu_\tau$, the 100 MeV-mass right-handed neutrinos and the scalar with $m_\phi = 10\;$MeV.
For this figure we included two right-handed neutrinos as this is the minimum necessary to explain neutrino masses, although depending on their hierarchy there could be effectively one or three; the impact on the temperature of the CCSN remain largely unchanged as the number of right-handed neutrinos is changed.
The final temperature $T_\mathrm{BSM}$ achieved in this process depends on the initial SM CCSN neutrino temperatures $T_\mathrm{SM}$ and chemical potential of $\nu_e$ (the two other flavors of neutrinos and antineutrinos are expected to have negligible chemical potentials and $\mu_{\bar\nu_e}=-\mu_{\nu_e}$). We depict the representative value of $\mu_{\nu_e}$ and $T_\mathrm{SM}$ in the SM CCSN core after the bounce with a black star, see, \emph{e.g.}, Fig.~5 in Ref.~\cite{Raffelt:2025wty}.
If $\phi$ is present, it will interact with neutrinos modifying the SM CCSN temperature to a new $T_{\rm BSM}$ and drive the $\mu_{\nu_e}$ to zero.
The final temperature reached after the equilibration of $\phi$ with all six neutrino flavors and sterile states for such a case is $\sim2.5-3$ times larger\footnote{This increase in temperature, and resultant phenomenology, can also arise in other situations such as quark-hadron phase transitions which may happen within the SM, see e.g.~\cite{Fischer:2017lag,Pitik:2022jjh}.}.
However, if $\phi$ transitions into the false vacuum, it will lower the final equilibration temperature due to cost of pumping energy into the $\phi$ potential to push the field into the symmetric phase.
In addition, the neutronization burst, typically only detectable in the inverted mass ordering in electron neutrinos, would potentially become present in all six flavors and thus detectable at all detectors (e.g.~SuperK/HyperK/JUNO) instead of only Ar detectors (e.g.~DUNE){\footnote{Electron neutrinos from galactic CCSN can also be detected via charged-current interactions with oxygen in water Cherenkov detectors.}}.

It is clear that the LNV neutrino-self interactions should significantly affect the standard CCSN evolution. If the scenario becomes efficient after the core bounce, we expect drastic changes in the temperature, neutrino distributions, and electron fraction. Contrary to the scenarios affecting the infall-phase~\cite{Fuller:1988ega, Suliga:2024nng}, we do not expect the change in the pressure support, as it is already coming from the degenerate nucleon gas.
However, interestingly, looking at the results from a CCSN simulation including a proxy reaction for LNV neutrino self-interaction $\nu_e + N \rightarrow \bar\nu_e + N$ ~\cite{Rampp:2002kn}, even though the temperature of the core increases and electron fraction decreases, the bounce still takes place and changes to shock evolution are minimal. Ref.~\cite{Rampp:2002kn} also does not find significant changes to the neutrino heating.
Thus we anticipate that, while this scenario modifies a number of internal components of a CCSN, these changes should not alter the fact that neutrinos drive successful CCSN explosions.
To fully assess the impact of such drastic alterations on the shock-wave evolution, neutrino heating, and nucleosynthesis, a dedicated full CCSN simulation incorporating treatment of high temperature and low electron fraction nuclear matter should be performed, which is outside the scope of this work.

The second effect that we will consider is the possible restoration of the false vacuum, corresponding to the vanishing of $\vev{\phi}$, as the CCSN heats up to its maximum temperature.  Here, we note that unlike the case of the early Universe, we do not expect a uniform temperature across the CCSN.
Eventually, the regions of the star that have transitioned into the false vacuum would transition back to the true vacuum, which is the counterpart of the  cosmological event that presumably led to the nanohertz GWs motivating our work.  Here, once the field tunnels back into the true vacuum, it would oscillate around its new vev.  These oscillations will be damped by the decay of the scalar (into active neutrinos in this case), returning the thermal energy stored in the potential $V(\phi)$ back to the plasma akin to the reheating phase of inflation.
One might expect that this would generally lead to a temporally distinct nearly monochromatic neutrino flux with energy $\approx m_\phi/2$.
We note, however, that the regions of the CCSN where this happens will likely be within the neutrinosphere and thus the neutrinos will rethermalize, washing out this signal. 

\begin{figure*}[t]
\centering
\includegraphics[width=0.49\textwidth]{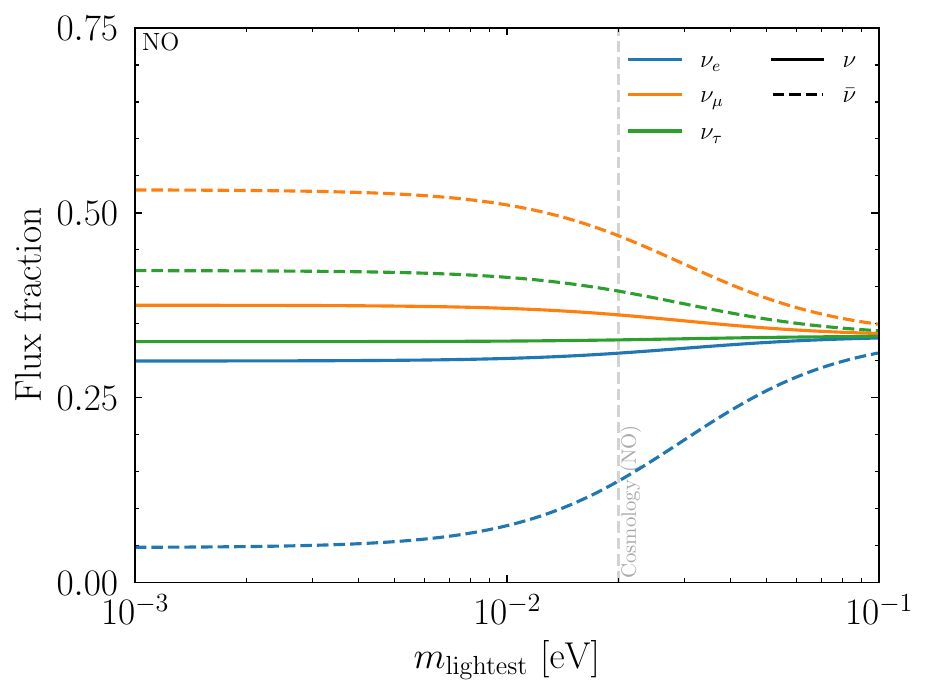}
\includegraphics[width=0.49\textwidth]{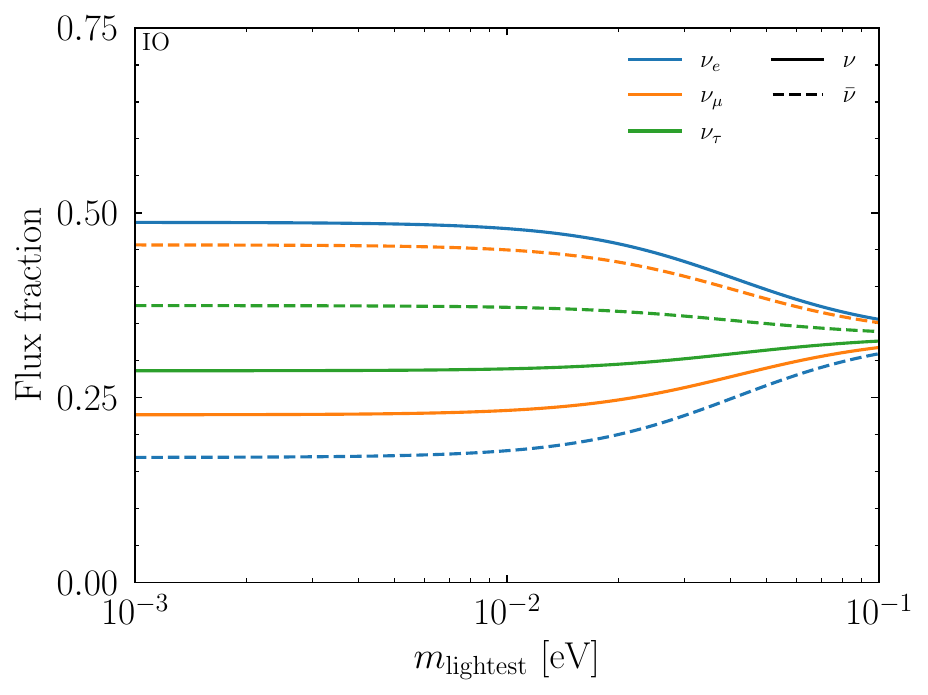}
\caption{The expected flux fraction, or flavor ratio, of any of the neutrino signals from the $\phi$ field at the Earth after undergoing the MSW effect.
In addition, the $\nu/\bar\nu$ ratio should be one.
The left panel shows the normal ordering (NO) and the constraint from cosmology \cite{Planck:2018vyg} in gray disfavoring masses to the right.  The right panel shows the inverted ordering (IO) which is nominally disfavored by cosmology while oscillation data remain unclear.}
\label{fig:flavor ratio}
\end{figure*}

In any scenario where either on-shell $\phi$'s or the $\phi$ condensate decays to neutrinos, the flux of neutrino mass eigenstate $i$ is expected to be $\propto m_i^2$ as $\phi$ generates neutrino masses.
Thus, the flux at the Earth of flavor $\nu_\alpha$ is $\Phi_{\nu_\alpha}\propto m_i^2P_{i\alpha}$ where the transition probability from production inside the CCSN to the Earth is given by
\begin{equation}
P_{i\alpha}=\sum_j|V_{ij}|^2|U_{\alpha j}|^2\,,
\label{eq:Pia}
\end{equation}
where $U$ is the PMNS matrix \cite{Pontecorvo:1957cp,Maki:1962mu}.
$V$ is the unitary matrix that diagonalizes the neutrino Hamiltonian at the production point in the mass basis:
\begin{equation}
H^m=\frac1{2E}\left[
\begin{pmatrix}
0\\&\Delta m^2_{21}\\&&\Delta m^2_{31}
\end{pmatrix}
+U^\dagger
\begin{pmatrix}
a\\&0\\&&0
\end{pmatrix}U\right]\,,
\end{equation}
by $VH^mV^\dagger=\Lambda$.
Here $\Lambda$ is the diagonal matrix of eigenvalues and $a\equiv 2E_\nu V_{\rm CC}$ with $V_{\rm CC}$ the charged-current matter potential.
The parameter $a$ quantifies the size of the matter effect and is large where the neutrinos from $\phi$ are produced, {\it i.e.}~$a\gg\Delta m^2_{31}$ where $|\Delta m^2_{31}|\sim2.5\times10^{-3}$ eV$^2$ is the atmospheric mass splitting.
Baked into equation \eq{eq:Pia} is the MSW \cite{Wolfenstein:1977ue,Mikheyev:1985zog} effect which adiabatically takes neutrinos from the matter basis to mass basis with no oscillations or jumps \cite{Mikheyev:1985zog,Parke:1986jy} en route.
Thus, under the adiabatic assumption, the flavor ratios do not depend on the neutrino energy.
To summarize, neutrinos are produced in the mass basis, they then propagate in the matter basis out of the star, traveling (as they decohere) to the Earth, where they are eventually  detected with a given flavor.

The expected flavor ratios of neutrinos coming directly from $\phi$ decay for both mass orderings as a function of the absolute neutrino mass scale are shown in Fig.~\ref{fig:flavor ratio} with the total $\nu/\bar\nu$ ratio exactly one, which is a unique and powerful signal.
In addition, it turns out that in the NO the antineutrino flavor ratio is outside the allowed region for any standard production in a CCSN \cite{Capanema:2024hdm}, as is the neutrino flavor ratio in the IO.
Thus there is a predictive signal in the flavor ratio for both mass orderings along with the neutrino-antineutrino equivalence.
In all cases it leads to fairly precise predictions for the flavor ratios of the extra neutrino signals that are testable in future Galactic CCSN observation.

\prlsection{Other Possible Effects}{.} 
The new sterile neutrino will mix with active neutrinos at the level of $(y\vev{\phi}/m_R)^2\sim10^{-9}$. This is currently allowed by terrestrial searches \cite{Fernandez-Martinez:2023phj} such as PIENU \cite{PIENU:2017wbj} but should be thoroughly tested by the proposed PIONEER experiment \cite{PIONEER:2022yag}\footnote{The DUNE near detector will likely improve upon PIENU's bounds, but may or may not reach the $10^{-9}$ level \cite{Ballett:2019bgd,Berryman:2019dme,Coloma:2020lgy,Breitbach:2021gvv,Moghaddam:2022tac,DUNE:2024wvj}.
In addition, since $\nu_R$ decays fast through $\phi$ channels to active neutrinos, we evade various supernova~\cite{Dolgov:2000jw, Fuller:2008erj, Rembiasz:2018lok, Mastrototaro:2019vug, Chauhan:2023sci, Carenza:2023old, Chauhan:2025mnn, DelaTorreLuque:2024zsr} and BBN bounds~\cite{Fuller:2011qy, Gelmini:2020ekg, Sabti:2020yrt, Boyarsky:2020dzc, Mastrototaro:2021wzl, Fuller:2024noz, Dev:2025pru}.}.

It is clear that such a model modifies the neutrino signal from a Galactic CCSN, although not at a level relevant for the observation of SN1987A given the very low statistics.
Unlike in the standard CCSN, where the electron neutrinos and antineutrinos diffusing from the neutrinosphere behave differently and their fluxes are dissimilar, after the equilibration between neutrinos and $\phi$, the electron neutrino and antineutrino fluxes will be similar.
This could have consequences for the supernova nucleosynthesis, as the differences in electron neutrino and antineutrino fluxes affect the amount of proton and neutron rich material ejected from the CCSN core~\cite{Qian:1996xt}. In addition, the expected drop in the core's electron fraction could also lead to increased amount of neutron rich ejecta potentially enabling at least partial r-process nucleosynthesis. However, it is challenging to decisively say how such a change that affects the core of the star will modify the amount of neutron rich ejecta without a full dedicated CCSN and nucleosynthesis simulations.

Such a scalar will also modify the signal of the diffuse supernova neutrino background (DSNB)~\cite{https://doi.org/10.1111/j.1749-6632.1984.tb23362.x, Krauss:1983zn, Beacom:2010kk, Lunardini:2010ab, Kresse:2020nto, 2023PJAB...99..460A, Suliga:2022ica}.
Super-Kamiokande (SK) with gadolinium may well be close to a detection \cite{Super-Kamiokande:2021jaq, harada_2024_12726429} and further data from SK and others may well achieve detection in coming years~\cite{JUNO:2022lpc}.
While considerable astrophysical uncertainties persist~\cite{Moller:2018kpn}, it is conceivable that future measurements and calculations could reduce these uncertainties.
This, combined with a high significance detection, which may require many years of Hyper-Kamiokande \cite{Hyper-Kamiokande:2018ofw} with gadolinium, may open up the DSNB as another probe of our scenario.

It may also be the case that the FOPT in the early Universe at $\ord{\rm 10~MeV}$  temperatures led to the production of primordial black holes (PBHs), see {\it e.g.} Refs~\cite{Jedamzik:1998hc, Byrnes:2018clq, Davoudiasl:2019ugw, Davoudiasl:2021ijv, Carr:2020gox, Ashoorioon:2022raz, Bhaumik:2025vlb}. These temperatures can lead to masses of $\gtrsim30$ $M_\odot$ depending on the details of the FOPT.
PBHs at $\sim35$ $M_\odot$ may correspond to a marginal excess in the spectrum of BH masses seen by LIGO-VIRGO-KAGRA \cite{LIGOScientific:2025pvj}.

\prlsection{Summary}{.}
In this letter, we have considered a simple currently unconstrained neutrinophilic scalar scenario that connects the signal at NANOGrav, neutrino masses, and potential signatures visible during a Galactic CCSN. 
We review the components of the scenario here:
\begin{enumerate}

\item NANOGrav has evidence for a gravitational wave signal consistent with an early Universe FOPT at temperatures $\mathcal{O}(10-100)$~MeV.

\item This phase transition may well be sourced by a scalar that is thermalized with the SM.

\item Neutrinos provide the least constrained means of thermalizing the scalar in the early Universe, provided that there is an additional right handed neutrino.

\item The right-handed neutrino provides for a low-scale seesaw mass generation mechanism for the active neutrinos which, in turn, sets the mass scale of the right-handed neutrinos.

\item  Conditions similar to that of the FOPT in the early Universe are realized inside a CCSN which can populate the scalar field.
 
\item The new lepton number violating fields modify the neutrino signal from a CCSN, providing unique signatures in time, energy, and the flavor ratio of neutrinos.

\end{enumerate}

Thus the neutrino signal from a future Galactic CCSN can confirm, or rule out, the minimal setup of a scalar field coupled to neutrinos potentially explaining both the pulsar timing array data from NANOGrav and others, as well as neutrino masses, in much of the interesting parameter space.

\begin{acknowledgments}
\prlsection{Acknowledgements}{.} The work of H.D.~and P.B.D. is supported by the US Department of Energy under Grant Contract DE-SC0012704.
The work of A.M.S.~is supported by the Department of Energy grant No.~DE-AC02-07CHI11359:~\emph{ Neutrino Theory Network Program}.
\end{acknowledgments}

Data is available in the arXiv ancillary files.

\bibliography{PGW-Supernova}

\end{document}